\documentclass[draft,showpacs,preprintnumbers,amsmath,amssymb]{revtex4}
\usepackage{graphicx}% Include figure files
\usepackage{dcolumn}% Align table columns on decimal point
\usepackage{bm}% bold math
\usepackage{latexsym}
\begin{document}
\author{\"{O}zg\"{u}r Delice}
 \email{odelice@boun.edu.tr}
 \affiliation{Department of Physics, Bo\u{g}azi\c{c}i University, 34342 Bebek, Istanbul, Turkey}
\title{Local cosmic strings in Brans-Dicke theory with a cosmological constant}
\begin{abstract}
It is known that Vilenkin's phenomenological equation of state for
static straight cosmic strings is inconsistent with Brans-Dicke
theory. We will prove that, in the presence of a cosmological
constant, this equation of state is consistent with Brans-Dicke
theory. The general solution of the full nonlinear field
equations, representing the interior of a cosmic string with a
cosmological constant is also presented.
\end{abstract}
\maketitle
\section{Introduction}
Cosmic strings are linear topological defects which may have been
produced in the early universe during phase transitions in general
unified theories \cite{cosmicstring},\cite{Hindmarsh}. They were
believed to have important implications on  cosmology and
astrophysics such as their role in primordial density fluctuations
in the early universe where galaxies etc. form. Although
accumulation of observational data ruled out this possibility and
they cannot contribute more than $\sim 10\%$ \cite{wyman,bevis} to
the density fluctuations, the interest in cosmic strings was
renewed \cite{Kibble} due to several reasons, such as their
relations with fundamental strings in string theory
\cite{sarangi,Dvali,copeland}.

An infinitely long, straight, local gauge string is characterized
by i) a special form for the energy momentum tensor of the string
$T^{0}_{\phantom{a}0}=T^{z}_{\phantom{0}z}\neq0$ and others
vanishing; ii)a Lorentz boost invariance along the symmetry ($z$)
axis, due to Vilenkin \cite{Vilenkin}. The investigation of an
Abelian-Higgs type theory for a Nielsen-Olesen type $U(1)$ vortex
\cite{Garfinkle,Linetvortex} strengthens the above prescription
for a straight string. Exact interior thick string solutions
satisfying the above energy-momentum tensor have been found by
Gott \cite{Gott}, Hiscock \cite{Hiscock}, and Linet
\cite{linetstr}.

The Brans-Dicke (BD) theory \cite{brans-dicke} is a natural
generalization of Einstein's general relativity (GR) where the
gravitational coupling constant is replaced by a scalar field.
This theory has some desirable properties; for instance, it obeys
the weak equivalence principle and it is also compatible with the
solar system experiments for certain ranges of its parameter
\cite{Faraoni}. It has been applied to several issues ranging from
inflation schemes in cosmology to quantum gravity. The inclusion
of scalar fields in GR is also suggested by different theories,
for example, as a  dilaton field which naturally arises in the low
energy limit of string theories or in the Kaluza-Klein theories
\cite{brans}.

In the framework of scalar-tensor theories, cosmic strings are
investigated in detail \cite{GundlachOrtiz}-\cite{Sen}. Sen {\it
et al.} \cite{SenBB} and later Arazi {\it et al.} \cite{Arazi}
demonstrated that the BD field equations are inconsistent with a
local string having the above energy-momentum tensor. Moreover,
Gregory and Santos \cite{Gregory} studied the cosmic string as an
isolated self-gravitating Abelian-Higgs vortex in dilaton gravity,
and they showed that the above energy-momentum tensor results in a
contradiction in the field equations in general, and one must take
into the account the other stresses as nonvanishing. However,
their investigation also shows that if there exists a mass term in
the theory, than the contradiction is removed and Vilenkin's
prescription might be appropriate. The inconsistency is also
removed if we have a more general scalar-tensor theory
\cite{SenB}, or the string is nonstatic \cite{Sen}. In this
report, I will show that, in the presence of a cosmological
constant, Vilenkin's prescription is consistent with BD theory. I
also present some exact solutions of the BD field equations
satisfying the above energy-momentum tensor and representing an
infinitely long, static straight, local gauge cosmic string.
\section{Field Equations}
The Brans-Dicke theory is described, in the presence of a
cosmological constant, by the action:
\begin{eqnarray}\label{BDaction}
S&=&\int d^4x
\sqrt{-g}\left\{\phi(-R+2\Lambda)+\frac{\omega}{\phi}g^{\mu\nu}\partial_\mu\phi
\partial_\nu \phi\right\}%\nonumber\\&&
+  S_m[\Psi,g];
\end{eqnarray}
here $R$ is the Ricci scalar, $\Lambda$ is cosmological constant,
$g$ is the determinant of the spacetime metric $g_{\mu\nu}$,
$\phi$ is the Brans-Dicke scalar field, $\omega$ is the
Brans-Dicke parameter, and $S_m$ denote the action of matter
fields $\Psi.$ We use units in which $c=\hbar=1$ and mostly plus
signature. The action (\ref{BDaction}) yields the field equations

\begin{eqnarray}
 \label{metric-eqn} G_{\mu\nu}+\Lambda
g_{\mu\nu}&=&\frac{T_{\mu\nu}}{\phi}+\frac{\omega}{\phi^2}\left(
\phi_{,\mu}
\phi_{,\nu}-\frac{1}{2}g_{\mu\nu}\phi^{,\alpha}\phi_{,\alpha}
\right)%\nonumber
%\\&&
+\frac{1}{\phi}\left(\phi_{,\mu;\nu}-g_{\mu\nu}\Box\phi
\right),\phantom{aa}\\
\label{scalar-eqn}(2\omega+3)\Box\phi&=&-2\Lambda
\phi+T^\mu_{\phantom{m}\mu}.
\end{eqnarray}
Here, $T_{\mu\nu}$ is the energy-momentum tensor of the matter
fields. Note that, in this frame (\ref{BDaction}), the energy
conservation equation
\begin{equation}\label{encons}
\nabla_\mu T^\mu_{\phantom{a}\nu}=0
\end{equation}
 holds.

Let us consider a general cylindrically symmetric, static metric
\begin{equation}
ds^2=e^{2(K-U)}(dr^2-dt^2)+e^{2U}dz^2+e^{-2U}W^2d\theta^2,
\end{equation}
where $t,r,z,\theta$ denote the time, the radial, the axial and
the angular cylindrical coordinates with the ranges
$-\infty<t,z<\infty,$ $0\le r <\infty$, $0\le \theta \le 2\pi$ and
the metric functions $K,U,W$ and the BD scalar field $\phi$ are
the functions of $r$ only. For the string, we consider Vilenkin's
prescription and take the only nonvanishing terms of
energy-momentum tensor of the form:
\begin{equation}\label{Tmnstring}
T^0_{\phantom{0}0}=T^z_{\phantom{z}z}=-\sigma(r). \
\end{equation}
 Then, the field equations are:
\begin{eqnarray}
&& -\frac{W''}{W}+K'\frac{W'}{W}-U'^2 =
\frac{\omega}{2}\left(\frac{\phi'}{\phi}\right)^2
-(K'-U')\frac{\phi'}{\phi} %\nonumber\\
%&&\phantom{aaaaaaaaa}
+ \frac{\phi^{''}}{\phi}+\frac{W'}{W}
\frac{\phi'}{\phi}
+(\Lambda+\frac{\sigma}{\phi})e^{2(K-U)},\label{Gtt}\\
\label{Grr} &&K'\frac{W'}{W}-U'^2 =
\frac{\omega}{2}\left(\frac{\phi'}{\phi}\right)^2-(K'-U')\frac{\phi'}{\phi}
%\nonumber\\
%&&\phantom{aaaaaaaaaa}
-\frac{W'}{W}\frac{\phi'}{\phi}
-\Lambda e^{2(K-U)}, \\%\ee
&&\frac{W''}{W}-2U''-2U'\frac{W'}{W}+K''+U'^2=-\frac{\omega}{2}\left(\frac{\phi'}{\phi}\right)^2
%\nonumber\\
%&&\phantom{aaaa}
-\frac{\phi^{''}}{\phi}-\frac{W'}{ W}\frac{\phi'}{\phi}%\nonumber\\
+U'\frac{\phi'}{\phi}
-(\Lambda+\frac{\sigma}{\phi})e^{2(K-U)},\label{Gzz}\\
&&\label{G33} K''+U'^2 =
-\frac{\omega}{2}\left(\frac{\phi'}{\phi}\right)^2 -
U'\frac{\phi'}{\phi}-\frac{\phi''}{\phi}%\nonumber\\
%&&\phantom{aaaaaaaaaaa}
-\Lambda e^{2(K-U)},\\ %\ee \be
\label{phieq} &&\phi''+\phi'\frac{W'}{W}=-\frac{2}{2\omega+3}(\Lambda\phi+\sigma)\,e^{2(K-U)}. %\ee
\end{eqnarray}
The energy conservation equation (\ref{encons}) yields
$K'\sigma=0$. Thus, in order to have a string ($\sigma\neq 0$), we
must have
\begin{equation}\label{K0}
K'=0,
\end{equation}
and hereafter we take $e^K=1.$ Using this and Eqs. (\ref{Grr}) and
(\ref{G33}), we find
\begin{equation}
\phi''+\phi'\frac{W'}{W}=-2\Lambda\phi e^{-2U}.
\end{equation}
Considering this equation with(\ref{phieq}) and taking into
account (\ref{K0}) we have:
\begin{equation}\label{sigma}
\sigma=2(\omega+1)\phi\Lambda.
\end{equation}
This expression says that, for $\Lambda=0$, we must have either
$\sigma=0$, i. e. no string at all, or $\omega\rightarrow \infty$,
if we have $\sigma\neq 0$. Thus, if we insist on $\sigma\neq 0$,
the solution reduces to the corresponding solution in GR. This
fact was noticed by \cite{SenBB} and they concluded that BD theory
is inconsistent for local strings, since for this case the only
solution is the same as GR one. However, for $\Lambda\neq
0,\omega\neq-1$ we have $\sigma\neq 0$, and for BD theory with a
cosmological constant, Vilenkin's prescription for local strings
is  consistent.
\section{Solutions}
Having proved that we can have a nontrivial string for BD theory
with a cosmological constant, let me introduce exact solutions
representing the interior regions of the string. Note that for
local gauge strings we must have boost invariance along $t,z$
directions and this requires $U=0$. Then, the field equations
reduce to the following set:
\begin{eqnarray}
&&\phi''+\phi'\frac{W'}{W}=-2\Lambda\phi,\label{cdfe1}\\
\label{cdfe3} &&
\frac{\omega}{2}\left(\frac{\phi'}{\phi}\right)^2=\frac{W'}{W}\frac{\phi'}{\phi}
+\Lambda, \\%
&&\frac{(\phi W)''}{\phi W}=-2\Lambda(2+\omega),\label{csfe4}
\end{eqnarray}
together with (\ref{sigma}). Now, we limit our analysis to
$\omega>-2$. (For $\omega<-2$, the results are the same as
$\Lambda\rightarrow -\Lambda$.) With this limitation, we have two
different sets of solutions depending on the sign of $\Lambda$.
\subsection{$\Lambda>0$}
For this case the Eq. (\ref{csfe4}) can be integrated to give
\begin{eqnarray}
W\phi&=&A \sin(\alpha\,r) +B\cos(\alpha \,r),\quad
\Lambda>0,\label{PW1}
\end{eqnarray}
where
\begin{equation}
\alpha=\sqrt{2\Lambda(2+\omega)}>0.
\end{equation}
 We choose $B=0$ due to cylindrical symmetry. By
considering Eqs. (\ref{cdfe1}) and (\ref{cdfe3}) we find
\begin{eqnarray}
W(r)&=&A \sin^{(\omega+1)/(\omega+2)}(\alpha\,r)\,\tan^{\epsilon/(\omega+2)}((\alpha/2)\,r),\\
\phi(r)&=&\sin^{1/(\omega+2)}(\alpha\,
r)\tan^{-\epsilon/(\omega+2)}((\alpha/2)\,r),
\end{eqnarray}
where $\epsilon=\pm1$. The metric has the form
\begin{equation}\label{strmetr}
ds^2=-dt^2+dr^2+dz^2+W^2d\theta^2.
\end{equation}
 Regularity conditions on the axis for
cylindrical symmetry require $W(0)=0$ and $W'(0)=1$. The first
one is automatically satisfied  and the second one is satisfied if
\begin{equation}
A=2^{-\omega/(2\omega+4)}/\sqrt{(\omega+2)\Lambda}.
\end{equation}
Hence the solution is regular and free of a conical singularity on
the axis. For $\epsilon=1$ it is also free of any singularity on
the axis since scalars constructed from curvature elements are
regular. For example, the Ricci scalar is
\begin{eqnarray}
R=\frac{2\Lambda\left[\omega+(\omega+1)^2(1+\cos(\alpha\,r))\right]}
{(\omega+2)\cos((\alpha/2)\,r)}\phantom{a}
\end{eqnarray}
and is regular on the axis. Other scalars have the similar
expressions. Thus, we see that the interior string solution is
smooth and regular near the axis. However, the solution is
singular at $r=r_c=\pi/(2\alpha),$ hence the radius of the string
should be smaller than $r_c.$

The mass per unit length of the string having radius $r_0$ can be
calculated using
\begin{equation}
m=\int_0^{2\pi}d\theta \int_0^{r_0}\sigma(r)W(r)dr.
\end{equation}
This yields
\begin{equation}
m=2\pi\frac{2^{\omega/{(\omega+2)}}(\omega+1)}{(\omega+2)}\sin^2\left((\alpha/2)r_0
\right).
\end{equation}
The solution with $\epsilon=-1$ must be excluded since it is
singular on the axis.
\subsection{$\Lambda<0$}
For this case, from (\ref{csfe4}) and cylindrical symmetry we must
have
\begin{equation}
W\phi=A\sinh{(\beta r)}
\end{equation}
with
\begin{equation}
\beta=\sqrt{|2\Lambda(\omega+2)|},
\end{equation}
 which yields, with the metric of the form (\ref{strmetr}),
\begin{eqnarray}
W(r)&=&C \sinh^{(\omega+1)/(\omega+2)}(\beta\,r)
\tanh^{\epsilon/(\omega+2)}((\beta/2)\,r),\\
\phi(r)&=&\sinh^{1/(\omega+2)}(\beta\, r)
\tanh^{-\epsilon/(\omega+2)}((\beta/2)\,r),
\end{eqnarray}
where $\epsilon=\pm1$. The analyses of the regularity conditions
and curvature scalars show that the solution with $\epsilon=1$ is
smooth, regular, and free of any singularity. Unlike $\Lambda>0$,
there is no upper limit on the value of the radius of the string.
The solution with $\epsilon=-1$ is singular on the axis and must
be excluded. The mass per unit length of this solution
($\epsilon=1$) with radius $r_0$ is
\begin{equation}
m=2\pi\frac{2^{\omega/{(\omega+2)}}(\omega+1)}{(\omega+2)}\sinh^2{\left((\beta/2)\,r_0\right)}.
\end{equation}
For all of these solutions $\sigma$ and $m$ are positive.
\section{Conclusions}
In this report we have proved that, unlike the case $\Lambda=0$
\cite{SenBB,Arazi}, in the presence of a cosmological constant,
$\Lambda$, the phenomenological equation of state for local cosmic
strings $T^0_{\phantom{a}0}=T^z_{\phantom{a}z}=-\sigma\
(\sigma>0)$ is consistent with Brans-Dicke theory (for $\omega\neq
-1,-2$). This result is in accordance with the analysis of Gregory
and Santos \cite{Gregory}. We have also presented exact solutions
for full Brans-Dicke field equations with a $\Lambda$ term for
this configuration and discussed some of their properties. Further
properties of these solutions with possible exterior solutions,
analogues of GR results \cite{Linet,tian},  are under
investigation \cite{Delice} and will be presented in elsewhere.

There is an interesting duality between GR and BD theories for
cosmic strings. In GR, Vilenkin's anzats is a good choice;
however, when there is a cosmological constant, one must take into
account the other stresses as nonvanishing, otherwise there is no
static solution\cite{tian}. In BD theory, however, as we have
shown, the situation is reversed and if there is no cosmological
term, then there is no solution, but if there is a cosmological
constant, then we can have strings with above properties in BD
theory.
%\begin{acknowledgments}
%I would like to thank Metin Ar\i k and Ali Kaya for reading the
%manuscript and useful discussions.
%\end{acknowledgments}

\end{document}